\documentclass{aastex}
\usepackage{spr-astr-addons_mod}
\usepackage{url}

\RequirePackage{color}

\begin{document}

\title{The Cosmic Origins Spectrograph:\\On-Orbit Instrument Performance}
\slugcomment{}
\shorttitle{COS On-Orbit Performance}
\shortauthors{Osterman et al.}

\author{S.\,Osterman,\,J.\,Green,\,C.\,Froning,\,S.\,B\'{e}land}\and
\author{E.\,Burgh,\,K.\,France,\,S.\,Penton}
   \affil{Center for Astrophysics and Space Astronomy, University of Colorado, Boulder, CO 80309, USA}
\and
\author{T.\,Delker,\,D.\,Ebbets} 
\affil{Ball Aerospace Technologies Corporation, Boulder, CO 80301, USA}
\and
\author{D.\,Sahnow} 
\affil{Department of Physics \& Astronomy, The Johns Hopkins University, Baltimore, MD 21218, USA }
\and
\author{J.\,Bacinski,\,R.\,Kimble}
   \affil{Goddard Space Flight Center, Greenbelt, MD 20771, USA}
\and
\author{J.\,Andrews,\,E.\,Wilkinson}
    \affil{Southwest Research Institute, Boulder, CO 80302, USA}
\and
\author{J.\,McPhate,\,O.\,Siegmund}
   \affil{Space Sciences Laboratory, University of California, Berkeley, CA 94720, USA}
\and
\author{T.\,Ake,\,A.\,Aloisi,\,C.\,Biagetti,\,R.\,Diaz,\,W.\,Dixon}\and
\author{S.\,Friedman,\,P.\,Ghavamian,\,P.\,Goudfrooij}\and
\author{G.\,Hartig,\,C.\,Keyes,\,D.\,Lennon,\,D.\,Massa}\and
\author{S.\,Niemi,\,C.\,Oliveira,\,R.\,Osten,\,C.\,Proffitt}\and
\author{T.\,Smith,\,D.\,Soderblom}
    \affil{Space Telescope Science Institute, Baltimore, MD 21218, USA}

%

\begin{abstract}
The Cosmic Origins Spectrograph (COS) was installed in the Hubble Space Telescope in May, 2009 as part of Servicing Mission 4 to provide high sensitivity, medium and low resolution spectroscopy at far- and near-ultraviolet wavelengths (FUV, NUV).  COS is the most sensitive FUV/NUV spectrograph flown to date, spanning the wavelength range from 900\AA~to 3200\AA~with peak effective area approaching 3000\,cm$^{2}$.  This paper describes instrument design, the results of the Servicing Mission Orbital Verification (SMOV), and the ongoing performance monitoring program.

\end{abstract}


\section{Introduction}

The Cosmic Origins Spectrograph (COS) is a fifth generation Hubble Space  Telescope (HST) science instrument providing high sensitivity, low and moderate resolution FUV and NUV spectroscopy between 900 and 3200\AA~(\citet{Green2003}).  COS is optimized to provide high sensitivity, moderate resolution spectroscopy of point-like objects with the principal goal of using distant quasi-stellar objects and faint stars to probe the density, temperature and composition of the intervening intergalactic and interstellar medium, exploring the web of dark baryonic mater in the universe.  With a sensitivity 10-30 times higher than any previous FUV instrument and with extremely low background, COS increases observational efficiency by a factor of 100 when viewing faint targets and in the first year on orbit has increased the integrated UV absorption line-of-sight ($\Delta$z) observed by a factor of 20.  For a comprehensive discussion of COS science goals and accomplishments see the article by C. Froning in this volume.

\subsection{Instrument Description}

\begin{deluxetable}{cccc}  
\tabletypesize{\scriptsize}
\tablecaption{COS Observing Modes\label{COSSpecModes}}
\tablewidth{0pt}
\tablehead{
%
\colhead{Channel} & \colhead{Wavelenth} & \colhead{Resolving Power} & \colhead{{V$_{\text{mag}}$\tablenotemark{a}} } \\

 \colhead{} & \colhead{Range (\AA)} & \colhead{$(\lambda/\Delta\lambda)$} & \colhead{}
}
\startdata 
 G130M & 
{900-1450~\tablenotemark{b}}
 					       & {16,000-21,000~\tablenotemark{c}} 
					       & 19.1 [{\it 16.3}] \\

 G160M & 1405-1775  & 16,000-21,000~\tablenotemark{c} & 16.7 [{\it 14.1}] \\
G140L   & $<$900-2050    & 1500-4000~\tablenotemark{c}        & 20.6  [{\it 17.8}]\\
G185M  & 1700-2100  & 16,000\,-$>$20,000 & 17.8  [{\it 15.7}]\\
G225M  & 2100-2500  & $>$20,000	       & 17.5  [{\it 15.8}]\\
G285M  & 2500-3200  & $>$20,000	       & 16.9  [{\it 15.6}]\\
G240L   & 1650-3200  & 2100-3900        & 20.4  [{\it 18.7}]\\
MIRRORA        & 1650\,-$>$3200  & 0.05\arcsec\ Imaging   &  \\
\enddata

\tablenotetext{a}{~S/N-10 per resolution element in 10Ksec for a B5V star at the nominal central wavelength for each mode using Castelli-Kurucz model [{\it or a flat (F$\lambda$) continuum}].  Values derived with the COS exposure time calculator based on mid 2010 calibration values.}
\tablenotetext{b}{~G130M modes extending to 900\AA~with resolution comparable to G140L will be available for cycle 19.}
\tablenotetext{c}{~Upper limits to FUV mode resolving powers are based on line spread function models and have not been verified on orbit and apply to $\lambda  >$\,1130\AA~only.}
\end{deluxetable}
\begin{figure*}[t]
\begin{center}
\includegraphics[width=0.85\textwidth]{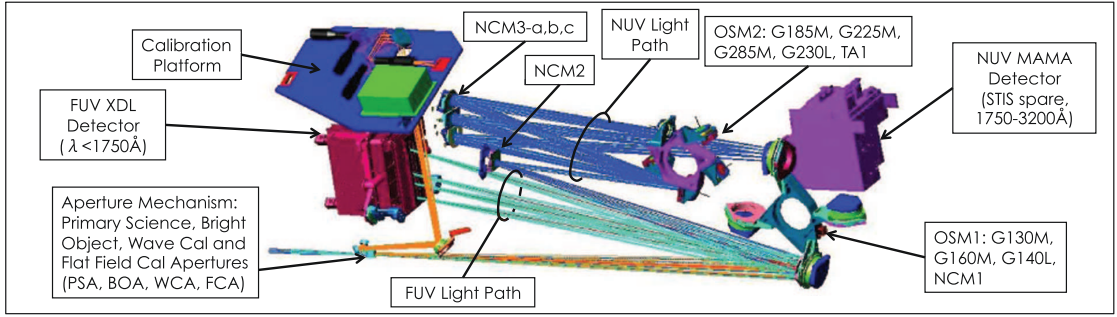}
\end{center}
\caption{COS Light Path.  The FUV channel is a modified Rowland circle spectrograph with three selectable gratings and is distinguished by having only one reflection after the Hubble OTA.  The holographically ruled, aspheric FUV diffraction gratings provide dispersion, reimaging, astigmatism control and aberration correction in a single reflection.   The NUV channel (four gratings and a mirror) is a modified Czerny Turner Spectrograph with the spectra reformed into three stripes on the NUV MAMA detector.}
\label{fig:COS_Op_Path}
\end{figure*}

The COS instrument is described in detail in Green (2003) and in the COS instrument handbook
(Dixon and Nemi (2003)), 
and is briefly reviewed here.  COS has two channels; a far-ultraviolet (FUV) channel, nominally providing coverage from 1150 to 1775\AA , but with coverage extended after launch to 900\AA, and a near-ultraviolet (NUV) channel operating from 1750 to 3200\AA.   The FUV band is covered by two medium resolution diffraction gratings (G130M, G160M)  and one low resolution grating (G140L).\footnote
{\,We use the HST nomenclature for grating mode identification: G{\it nR} describes the grating with nominal wavelength near {\it n}\,nm and spectroscopic resolution {\it R} (medium (M) or low (L)).  Central wavelength setting {\it m} is called out as G{\it nR-m} with {\it m}  in \AA.   E.g. the nominal (1300\AA) wavelength setting of the G130M grating is referred to as `G130M-1300.'} 
       NUV coverage is provided by three medium resolution gratings (G185M, G225M and G285M) and one broad band, low resolution grating (G230L), as well as a broad band, narrow field imaging mode (MIRRORA/MIRRORB)  intended primarily for target acquisition (TA) but capable of providing high resolution NUV imaging.  All spectroscopic modes have several supported central wavelength settings.  The optical paths are shown in figure \ref{fig:COS_Op_Path}.

COS is a slitless spectrograph; light from the Hubble Optical Telescope Assembly (OTA) enters COS through one of two selectable 2.5\arcsec\,diameter apertures admitting $>$95\% of the light from a point-like source. The primary science aperture (PSA) is windowless, so any short wavelength ($<$1150\AA) light remaining after the two reflections in the OTA will travel unobstructed to the COS optics.  Alternatively, observers may select the Bright Object Aperture (BOA), has a neutral density filter that attenuates light by approximately 150$\times$, although with reduced spectroscopic resolution.

%

Incoming light strikes one of the four optics on the first rotary optics select mechanism (OSM1).  The selected optic either diffracts light (G130M, G160M or G140L) onto the FUV detector or directs light into the NUV channel via the aspheric NCM1 mirror.

The FUV channel is a modified Rowland circle spectrograph; the diffraction gratings are holographically ruled onto aspheric concave substrates, providing aberration and astigmatism control, and the detector is a two segment cross delay line (XDL) device.   In addition to optic selection, OSM1 can also commanded to intermediate positions for the different central wavelengths associated with each grating to ensure continuous wavelength coverage despite the inter-segment detector gap.  OSM1 is also used to step the grating through four smaller fixed pattern positions (FP-POS) moves for each central wavelength to reduce sensitivity to fixed pattern noise.

The NUV channel is a modified Czerny Turner spectrograph with a two mirror collimator (NCM1 on OSM1 and the fixed NCM2), four flat diffraction gratings (G185M, G225M, G285M and G230L) and a flat mirror (TA1) on the second optics select mechanism, OSM2, and three camera optics (NCM3-a,b,c).  The camera optics stack three non-contiguous portions of the spectrum (stripes a, b,c) onto the NUV MAMA (Multi-Anode Microchanel Array, \citet{Argabright1998}) detector to make best use of the square detector format.  The TA1 mirror (MIRRORA mode) provides field of view  imaging (1\arcsec~unvignetted field of view) and supports target acquisition.  The TA1 mirror and the G285M and G230L gratings have fused silica windows which function as order-sorters, and, in the case of TA1, as a second, low efficiency ($\sim$5\%) mirror (MIRRORB mode).  COS observing modes are summarized in table \ref{COSSpecModes}. Like OSM1, OSM2 is typically stepped through FP-POS moves.

Both the NUV and FUV detectors can provide data as either time-tagged photon lists (TIME-TAG mode) or as accumulated images (ACCUM) without time-tag information on individual photon events.  Time-tagged observations allow for correction of spectral drift due to mechanism settling as well as other transient effects.  COS has an onboard calibration platform that can deliver wavelength calibration spectra (from one of two Pt/Ne emission line lamps) and a quasi-continuum spectra (from one of two D$_{2}$ lamps), illuminating the wavelength calibration aperture (WCA) and flat field aperture (FCA), respectively.  The wavelength calibration lamps provide simultaneous wavelength reference spectra for the grating modes, offset in the cross dispersion direction from the science spectra (at wavelengths longer than the lamp window cutoff at 1150\AA)
and a fiducial image in the MIRRORA/MIRRORB modes.  All apertures are on a common, movable aperture plate, so that the offset between the WCA and the center of the PSA and the BOA are fixed.  The calibration spectra allow registration of the spectra after grating moves, permit tracking of grating drift during an observation through a series of calibration lamp flashes during the exposure, and assist in target acquisition. 

\section{FUV Channel Performance}

\subsection{Detector Performance}
The FUV detector was designed and built by the Experimental Astrophysics Group at the University of California, Berkeley (\citet{Vallerga2002}, \citet{McPhate2010}).  The detector is a windowless, two segment XDL microchannel plate (MCP) device comprised of two 10$\times$85mm segments placed end to end with a 9mm gap (segments A and B).  The MCPs are coated with an opaque CsI photocathode for high efficiency at FUV wavelengths and are curved to approximate the COS FUV focal surface.  The detector active areas are digitized to 16384$\times$1024 ÔpixelsÕ each and provide 25-30$\mu$m resolution in the dispersion direction.  Pixels are approximately 6$\mu$m wide in dispersion by 24$\mu$m tall.  The COS FUV detector is a photon counting device, and can provide data either as a time-tagged photon list with with 32ms time resolution and pulse height information or as accumulated images (TIME-TAG or ACCUM modes).  It is important to note that the XDL detector does not posses  physical pixels in the sense of CCD pixels; the XDL electronics generate analog positions which are then digitized, creating the reported raw pixel values.  These then require geometrical and thermal correction prior to being converted to a physical location on the detector (\citet{Wilkinson2003})

The on-orbit background rate for the FUV detector is approximately 1 count per  resolution element (36x240 $\mu$m) per 10$^{4}$ seconds, although this increases  near the South Atlantic Anomaly (SAA).\footnote{\,Resolution elements (`resels') are used for both the NUV and FUV detectors to avoid confusion with the physical pixels encountered with CCD detectors.  An FUV resel is 6 pixels wide by 10 pixels height (36$\times$240$\mu$m) and an NUV resel is 3$\times$3 pixels (75$\mu$m square).}
    Both COS detectors are commanded to reduced voltage during SAA passages to prevent detector overlight.  The FUV detector background is largely uniform except for several weak features on segment B,  corresponding to low-pulse height events which can be eliminated by proper pulse height threshold selection (\citet{France2010}).

The FUV detector experienced a $\sim$20\% increase in MCP gain after launch compared to preflight testing (gain is the average number of electrons generated by the MCP stack per photon, and is related to but  not equivalent to the detector quantum efficiency).  This gain increase is attributed to gas absorption during the final 10 days prior to launch when the detector pumping was shut down.  Detector high voltages settings were decreased after launch to return the gain to the pre-launch levels.  During the first year on orbit gain has dropped by $>$\,30\% over most of the illuminated portion of detector, with greater decreases where geocoronal Ly-$\alpha$ falls, resulting in localized sensitivity decreases.  Gain sag was anticipated and can be corrected by increasing the detector voltage, by adjusting the pulse height filtering in the COS calibration pipeline (CALCOS), or by moving the spectrum in the cross dispersion direction to a ``fresh" portion of the detector.  Impact on science spectra can also be mitigated by flagging pixels with the most severe gain sag.  Gain sag also has implications for target acquisition (section 4).  Independent of gain sag, the detector sensitivity appears to be decreasing on-orbit faster than anticipated.  The cause for this is under investigation and is discussed in section 2.3.

\subsection{FUV Spectroscopic Performance}

COS is designed to provide medium and low resolution spectroscopy with better than 15~km\,s$^{-1}$ wavelength knowledge (150~km\,s$^{-1}$ in low resolution), with better than 8~km\,s$^{-1}$ typically achieved.  
Light from one of two Pt/Ne lamps (\citet{Nave2008}, \citet{Penton2008}) is reimaged onto the wavelength calibration aperture (WCA) on the aperture block and proceeds along either the FUV or NUV light path in parallel with the stellar spectrum.  The calibration spectrum is formed on the detector offset from the science spectrum in the cross dispersion direction, allowing registration of each spectrum and, if data is taken in time tag mode, correction of mechanism drift during an observation (\citet{ISR201009}).

\begin{figure}[ht]
\includegraphics[width=0.475\textwidth]{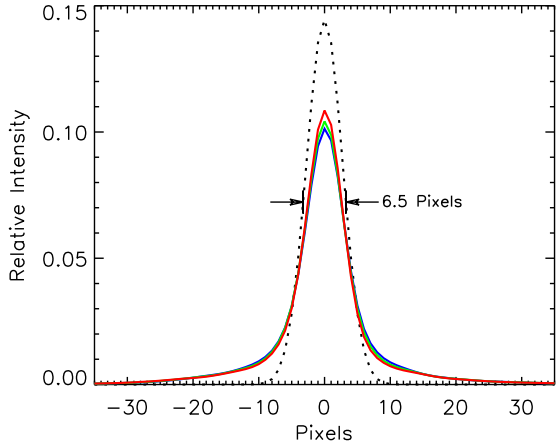}
\caption{Modeled line spread function for G130M at four wavelengths compared to the nominal 6.5 pixel FWHM Gaussian observed during ground testing. Red=1150\AA, green=1250\AA, blue=1350\AA, and dotted line shows the 6.5 pixel FWHM Gaussian profile.}
\label{fig:FUV_LSF}
\end{figure}

\begin{figure}[ht]
\includegraphics[width=0.475\textwidth]{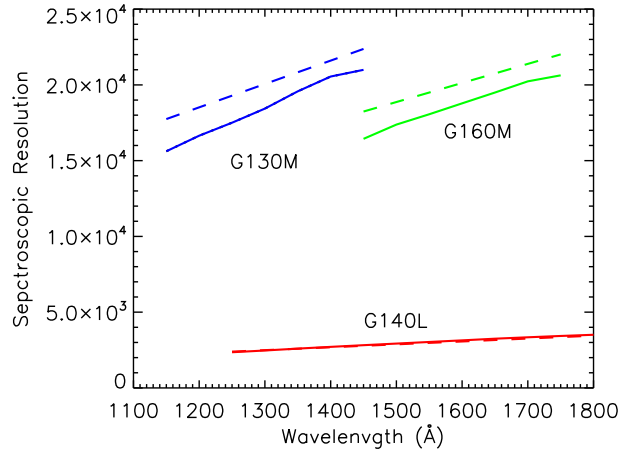}
\caption{ COS FUV spectroscopic resolution from modeled line spread function (solid) compared to the pre-launch nominal 6.5pixel FWHM (dashed).}
\label{fig:FUV_Res}
\end{figure}

COS was designed to provide spectral resolution $\lambda/\Delta\lambda$ $>$20,000 in the G130M and G160M channels.  While ground testing indicated that COS could support this resolution, mid frequency wavefront errors (MFWE) introduced by the Hubble OTA reduce COS spectral resolution by moving light from the line cores into the wings of the line spread function (LSF), resulting in full width at half maximum (FWHM) resolutions ranging from 16,000 to 21,000 for G130M and G160M, and Rayleigh criterion resolutions $>$\, 18,000 (table \ref{COSSpecModes} and figures \ref{fig:FUV_LSF} and \ref{fig:FUV_Res}).

An alternative expression of the instrument resolving power is the ability of COS to observe narrow FUV absorption features.  COS prelaunch performance estimates suggested that at 1300\AA~the G130M channel would be able to support 3$\sigma$ detection to a limiting equivalent width of W$_{\text{lim}}\sim12$m\AA~at S/N=10 per pixel for a line with Doppler parameter b=10 km\,s$^{-1}$, and W$_{\text{lim}}\sim33$m\AA~for much wider lines with b=100 km\,s$^{-1}$.  Modeling based on the broadened LSF predicts that the observed line spread function increases the limiting equivalent widths for these lines to $\sim$ 15m\AA~and 33.6m\AA, respectively.  Similar changes are seen in the G160M and G140L performance. (\citet{ISR200901})  Resolution in both the FUV and NUV channels is significantly reduced if the BOA is selected due to a wedge in the BOA filter, reducing M grating spectral resolution to 3500-5000.

The FUV channel is not intended to support cross dispersion imaging; rather, the holographic recording is intended to minimize image height to reduce detector areas.  The cross dispersion height for a point source ranges from $\sim$40 to 200 $\mu$m (0.35\arcsec\ to 1.8\arcsec)   depending on wavelength and grating (\citet{ISR201009}).  Gain sag in the FUV detector results in image displacement to lower cross dispersion values.  This has not been observed to introduce wavelength error.

\subsection{FUV Sensitivity}
On-orbit testing indicated that the prelaunch throughput requirements for COS were met or exceeded, with COS FUV medium resolution mode sensitivities ranging from 2 to more than 30 times higher than previous HST FUV capabilities. The low resolution G140L grating (as well as the G130M grating in new off nominal, low resolution configurations) are sensitive down to at least the Lyman limit (900\AA) with effective area comparable to or significantly larger than those of the Far Ultraviolet Spectroscopic Explorer (FUSE) over the same wavelength range, although at lower resolution (\citet{McCandliss2010}, \citet{Sahnow2000}).  

\begin{figure}[ht]
\includegraphics[width=0.475\textwidth]{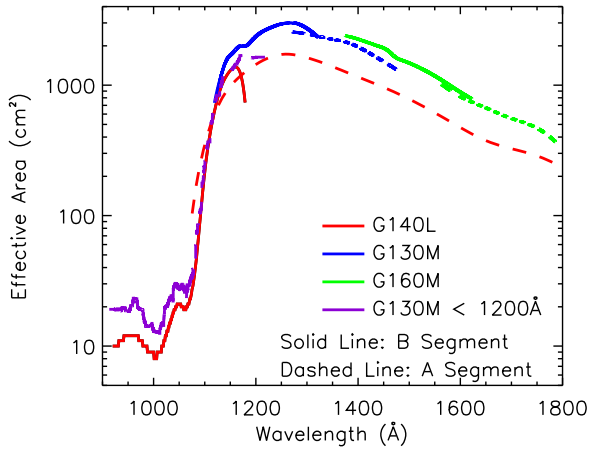}
\caption{COS FUV effective area including new short wavelength modes.  These curves represent throughput as of September, 2010.   Time dependent sensitivity is accommodated in the CALCOS pipeline}
\label{fig:FUV_A_Eff}
\end{figure}

The high sensitivity of the FUV channels is achieved by optimizing component efficiency, minimizing the number of reflections and by eliminating transmisive optics. The single FUV optic accomplishes diffraction and reimaging as well as aberration and astigmatism control.  The detector's CsI photocathode is optimized for the G130M and G160M bands and the detector is windowless to eliminate transmission losses.  This windowless design has the added advantage of enabling observations at wavelengths shorter than any previous HST instruments (section 2.4).  FUV sensitivity is summarized in table~\ref{COSSpecModes} and effective area is shown in figure ~\ref{fig:FUV_A_Eff}.

\begin{figure}[ht]
\includegraphics[width=0.475\textwidth]{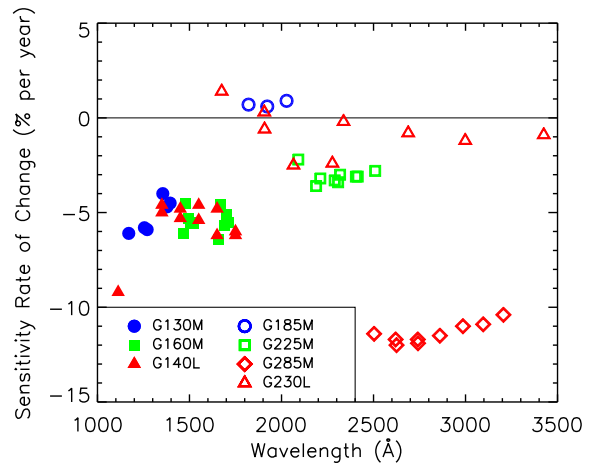}
\caption{COS sensitivity trends for February - August, 2010.  Solid symbols represent the FUV channel, open symbols represent the NUV.  Prior to this period the FUV gratings showed an increasing loss rate with wavelength ($>$10\%/year at 1700\AA), but this has disappeared in recent months. }
\label{fig:COS_TDS}
\end{figure}

The sensitivity of all FUV channels appears to be decreasing with time (figure~\ref{fig:COS_TDS} and \citet{ISR201015}).  Initially the loss of senditivity was most severe at longer wavelengths, but this wavelength dependence had disappeared by early 2010, and the rate of loss at all FUV wavelengths appears to be decreasing with time.


Hydrocarbon, contamination of optical surfaces, detector radiation damage, pipeline processing errors, and pointing errors leading to vignetting in the aperture were considered as possible causes of the drop in sensitivity and have been largely ruled out.  Another possibility, gain sag in the detector, was considered and discounted after demonstrating that the obvserved gain sag could not account for the observed loss of  detector performance.  Other pootential culprits -- geocoronal contamination of the calibration spectra and variability of the calibration target -- were likewise considered and discounted.

The remaining possible explanations focus on the detector photocathode.  The two mechanisms considered most likely  are contamination by water evolving from the composite structure and atomic oxygen encountered in low earth orbit.  The COS structure was kept under purge from late 2003 until launch in May, 2009, so it was expected to be extremely dry.  Based on photocathode testing by McPhate, model predictions of water desorption from the composite structure do not seem to provide sufficient moisture to account for the loss of detection efficiency. 

We are currently engaged in characterizing the sensitivity of CsI photocathodes to exposure to atomic oxygen, and preliminary results are suggestive that this could account for some or all of the photocathode degradation, although the investigation is ongoing and no final conclusion has made at the time of writing.  

\subsection{ FUV Sub-1150\AA~Performance}
The COS FUV channel was designed with a nominal wavelength range of 1150-1750\AA.  However, component level testing had shown that the G140L grating (the only optic in the G140L light path after the OTA) retained some efficiency down to at least 1066\AA~and that the windowless detector was sensitive  below 900\AA.  If the OTA retained some first surface reflectivity there was a good chance that COS would be able to provide spectra below 1150\AA\ on-orbit.  (\citet{Osterman2002}, \citet{McCandliss2010})

This was borne out early in SMOV, and subsequent analysis demonstrated sensitivity in the G140L mode on detector segment B ranging from $\sim$10\,cm$^{2}$ below 1050\AA~to $>$\;1000\,cm$^{2}$ at wavelengths above 1130\AA\;(fig. ~\ref{fig:FUV_A_Eff}), comparable the FUSE single channel effective area (10-20\,cm$^{2}$ per channel, \citet{Sahnow2000}).    In light of the short wavelength sensitivity of the G140L gratings two more modes were proposed and tested using the G130M grating.  These modes, G130M-1055 and -1096, place the G130M grating well outside of its nominal position, exchanging reduced spectral resolution (1500-2500) for moderate sensitivity ( $\sim$20\,cm$^{2}$) at wavelengths as short as 900\AA~(\citet{Osterman2010}).  In addition to having slightly higher sensitivity than the G140L modes, the G130M modes have narrower band pass.  As a result,  G130M-B segment wavelength coverage for the -1055 and -1096 settings ends before the grating and mirror throughput begins the rapid, two order of magnitude climb  between $\sim$1080 and 1200\AA, meaning that brighter targets, better suited to the relatively low sensitivity of the G130M-1055 and -1096 modes, can be observed without triggering the COS bright object protection limits if the A detector segment is disabled.

\subsection{FUV Signal-to-Noise and Detector Background}
COS achieves high signal-to-noise (S/N) observations through three means:  Maximizing efficiency, minimizing scattered light and detector background, and mitigating fixed pattern noise.  High efficiency is achieved through high detector efficiency, high reflectivity, high efficiency gratings and scrupulous contamination control on the ground and in material selection.  Low background is provided by the low noise detector (1 count per resel per 10$^{4}$ seconds for the FUV channel) and by controlling scattered light through the use of low scatter, ion etched diffraction gratings.  Fixed pattern noise control is achieved through routine use of FP-POS  moves, stepping the FUV spectrum through four $\sim$2.4\AA~offsets for each observation (19\AA~steps for G140L) .  

Flat fielding was intended to be provided by on-board flat field lamps and by observing relatively featureless stellar objects, but this has proven impractical for the FUV channel due to the complex interplay between the detector readout scheme and the MCP structure.  Mid-frequency terms such as the MCP moir\'{e} pattern, geometrical distortion and detector quantum efficiency enhancement grid wire shadows further complicate efforts to apply flat field correction.  The CALCOS data pipeline was intended to apply a flat field to the spectra, but it was found that adequate S/N could be achieved by flagging those areas subject to grid wire shadows, removing them from the concatenated spectra, and forgoing any other flat field correction (\citet{Danforth2010}).  Since no two FP-POS settings have the same set of pixels shadowed, grid wire tracking permits complete coverage of the spectral region with slightly reduced S/N in those areas that are masked out.  This method provides substantial improvement in achievable S/N; S/N of 19 may be obtained with no flat fielding and a single FP-POS settings compared to S/N of 51 being achieved in the FUV with four FP-POS settings and grid wire masking.  Masking is currently implemented in CALCOS using the bad pixel map.  Additional work is being performed on an iterative technique indicates that S/N$>$100 per resel may be possible for some FUV observations (\citet{Ake2010}).

\section{NUV Channel Performance}


\subsection{NUV Detector Performance}
The NUV MAMA detector has a 1024$\times$1024 pixel, 25.6mm square  anode with 25$\mu$m pixels and approximately 35$\mu$m resolution.  As with the FUV detector, data can be read out as either a TIME-TAG photon list or as an ACCUM image, although pulse height data is not available for either mode. COS MAMA on-orbit dark current has been slowly increasing since launch, from an initial level of  $\sim5\times10^{-4}$ cts/resel/sec  to the current rate of  $\sim3\times10^{-3}$cts/resel/sec.  The on-orbit background is presumably dominated by same phosphorescence in the MgF$_{2}$ window as observed with STIS, although at a much lower level (\citet{Kimble1997}).


\subsection{NUV Spectroscopic Performance}
\begin{figure}[ht]
\includegraphics[width=0.425\textwidth]{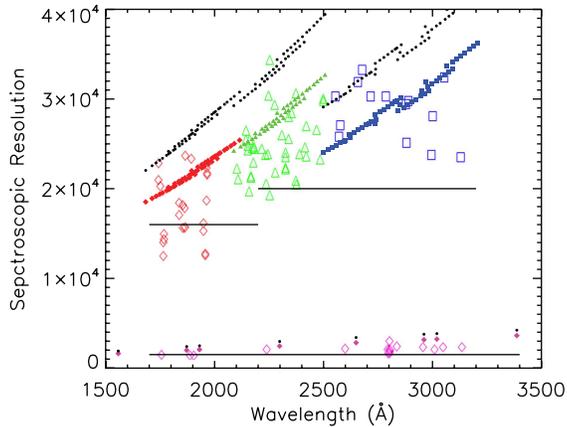}
\caption{On-orbit values for COS NUV Resolution (open symbols) compared to an assumed 2.4 pixel FWHM (solid symbols) and LSF model (black circles).  The lower than predicted performance is attributed to OTA breathing. G185M: Red diamonds.  G225M: Green triangles. G285M: Blue squares. G230L: Magenta diamonds.}
\label{fig:NUV_Res}
\end{figure}

The COS NUV channels exceed the pre-launch spectroscopic resolving power requirements. At NUV wavelengths, the LSF and cross dispersion profiles are dominated by focus offsets introduced by the OTA as a function of orbital phase (``breathing") and by the MAMA detector point spread function.  While a contributing effect in the NUV LSF,  the effect of the MFWE is of less significance at the longer NUV wavelengths than for the FUV (\citet{ISR201008}, \citet{ISR201010}).  NUV resolution is shown in in figure ~\ref{fig:NUV_Res}.

\subsection{NUV Sensitivity}
\begin{figure}[ht]
\includegraphics[width=0.475\textwidth]{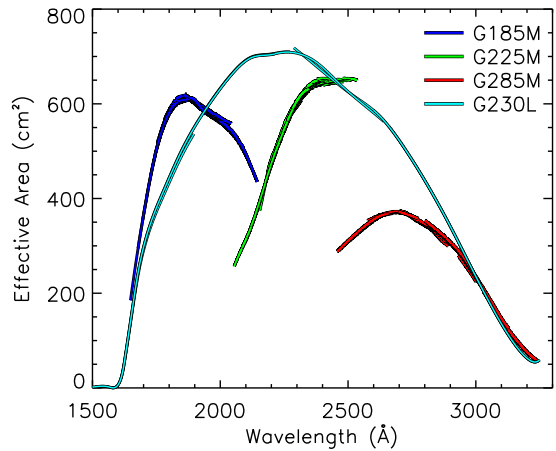}
\caption{COS NUV Effective Area as a function of wavelength (as of September, 2010).  NUV sensitivity is corrected in CALCOS for time dependent variation.}
\label{fig:NUV_A_Eff}
\end{figure}

COS NUV effective areas range from  $\sim$\,200-700\,cm$^{2}$ between 1650 and 3000\AA~(figure ~\ref{fig:NUV_A_Eff}).   With low-scatter optics and low detector background, this sets the limiting magnitude for a S/N=10 object as deep as V$_{\text{mag}}\sim$ 17.8 in $10^{4}$ seconds using the NUV M modes (table ~\ref{COSSpecModes}).

Two issues complicate COS NUV performance; the slow loss in efficiency apparent in two of the COS NUV gratings (G225M and G285M) (\citet{ISR201015}), and vignetting in the NUV spectra.  The NUV sensitivity loss is isolated to the G225M and G285M gratings and was first observed in ground testing.  The original NUV G185M, G225M and G285M gratings featured relatively high line densities (5870, 4800 and 4000 lines/mm, respectively) and operated close to the resonance regime with the wavelength to line spacing ratio $\lambda$/d $\sim$1.   The groove profiles were carefully designed to optimize performance in band, and vendor and GSFC  testing indicated that all of the gratings met the groove efficiency specification before final coating (gratings were delivered with a gold coating to facilitate testing).  However, when the gratings were coated with the NUV optimized MgF$_{2}$ protected aluminum, it became apparent that the grating efficiency was significantly suppressed.  Due to severe schedule constraints, the most expedient solution was to replace the G185M grating with the Al/MgF$_{2}$ coated G225M grating, trading slightly reduced spectral resolution (figure \ref{fig:NUV_Res}) for improved efficiency, and replacing the Al/MgF$_{2}$ coated G225M and G285M gratings with bare aluminum coated gratings.  These bare aluminum optics, while unacceptable for the G185M band due to the low reflectivity of oxidized aluminum at shorter wavelenths, achieved design performance in the G225M/G285M band.  

It was expected that the aluminum oxide layer would stabilize quickly, especially with the optics maintained in dry N$_{2}$ purge until launch (\citet{Bartl2004},~\citet{Herzig1993}).  Nevertheless, a grating monitoring program was established using the on-board Pt/Ne wavelength calibration lamps.  By comparing the measured intensities of lines that are observed by more than one grating, the relative efficiency of the NUV channels could be compared.   Assuming that the Al/MgF$_{2}$ coated optics were stable, it appeared that the bare aluminum optics were slowly losing efficiency, although witness mirror reflectivities appeared unchanged.  Absolute sensitivity measurements taken during the second thermal vacuum tests in 2006 confirmed this trend, and follow-on measurements indicated that one polarization of the incident light was being suppressed by the G225M and G285M gratings in band, possibly due to a continued growth of the Al$_{2}$O$_{3}$ oxide layer, or to a change in oxide structure.

This trend has continued on-orbit, with an annual sensitivity loss ranging from 3-4\%\;per year for the G225M modes to 10-12\%\;per year for G285M (figure ~\ref{fig:COS_TDS}.)  The time dependence of both the FUV and NUV sensitivity is being tracked on-orbit through a regularly scheduled sensitivity monitoring program and the COS data pipeline is updated to reflect these changes. 

After launch and final alignment a lower than expected efficiency was noted on one edge of all spectral stripes:  The count rate at the blue edge of the detector was $\sim$\;20\% lower than expected, ramping linearly to the expected value over the first $\sim$\;160 pixels.  This is believed to be due to vignetting of the beam at the NCM3 camera optics.  A correction is applied to the spectra in the data processing pipeline.

\subsection{NUV Signal-to-Noise}
The NUV channel of COS relies on the same approach to maximizing S/N as the FUV channel.  In the case of the NUV channel, flat fielding is significantly more straightforward but the detector background is substantially higher, as described above.  As with the FUV channels, FP-POS offsets are used to reduce fixed pattern noise, with 1.8-2.0\,\AA /step for the M gratings, and $\sim$19 \AA/step for the low resolution, G230L mode.  Comparison of flat field spectra obtained in ground testing and spectra obtained on-orbit indicates that there is no statistically significant difference and the two data sets were combined to produce a single deeper flat field.  
Using the CALCOS pipeline high S/N observations were obtained with the observed S/N as a function of counts closely following the ideal Poisson limit up to $\sim$S/N 50.  S/N$>$150 was demonstrated during the SMOV high S/N program. (\citet{ISR201003})

\subsection{NUV Imaging Mode}
OSM2 has a flat mirror (TA1) in addition to the four NUV gratings.  This mirror is intended primarily to support target acquisition (TA) by placing an image of the COS field of view on the NUV detector, but this mirror may also be used to obtain high angular resolution, narrow field of view NUV imaging (\citet{ISR201010}).

In the nominal mode (MIRRORA), the mirror places the target image near the center of the ÔbÕ stripe on the NUV detector.  The image FWHM varies as a function of HST OTA focus throughout the orbit, ranging from a FWHM of roughly 1.5 pixels to as much as 2.5 pixels (0.0235 arc sec/pixel plate scale).  The MIRRORA mode is sensitive from $\sim$1600\AA~to $\sim$3200\AA~with a small but measurable red leak extending to visible wavelengths. Two weak ghost images are located $\sim$20 pixels from the central peak at 0.1\% of the peak intensity. 

An alternate imaging mode (MIRRORB) uses first surface reflection off of the TA1 order-sorter to provide a reduced-sensitivity imaging mode ($\sim$20x reduction).  The finite thickness of the order-sorter window results in a secondary ÔghostÕ offset from the first image by 20 pixels.  The ghost image is taken into account in the TA centering algorithms.  Because MIRRORB employs a first surface reflection, this mode is sensitive to light down to the detector cutoff at $\sim$1150 \AA~for the primary image.
 

Finally, the MIRRORA imaging mode has been used to map out the aperture transmission function, showing 95\% of peak transmission out to $\pm$0.5\arcsec\ from the aperture center.

\section{Target Acquisition (TA)}
COS is a slitless spectrograph, relying on the superb pointing accuracy and stability of HST to achieve the desired wavelength accuracy and resolution.  To do this,  HST must place the target to within 0.1\arcsec\ of the center of the aperture for the FUV channel, and to within 0.04\arcsec\ for NUV in the dispersion direction. 0.3\arcsec\ centering is sufficient in the cross-dispersion direction to maintain resolution and wavelength precision, but 0.1\arcsec\ is desired. During SMOV and early Cycle 17  COS TA procedures were refined and were able to exceed  the required absolute wavelength precision for all three COS TA modes (NUV imaging, NUV spectroscopic, and FUV spectroscopic, see~\citet{ISR201014}).

Gain sag in the FUV channel results in a displacement of the image in the cross dispersion direction, complicating FUV spectroscopic target acquisitions.  In addition to approaches intended to mitigate gain sag in general, we are investigating the possibility of restricting the portion of the detector used for cross dispersion peakup to those areas least affected by gain sag.  The possibility of introducing pointing errors as a result of anomalously high detector background near the SAA has been addressed by ensuring that COS target acquisition exposures do not occur while HST is in or near the SAA. 

The  COS-to-fine guidance sensor offsets were updated in Cycle 17, further improving COS pointing accuracy. This has given us confidence in our ability to skip the relatively time-consuming pattern search (ACQ/SEARCH) procedure in cases where the target coordinates are well known.  NUV imaging TAs provide the best centering accuracy and are preferred over FUV spectroscopic TAs when absolute photometric accuracy is required.  NUV and FUV spectroscopic TAs are often faster than NUV imaging TAs and are often used when maximum S/N is desired.

\section{Conclusion}
COS is currently delivering unprecedented spectroscopic performance on a range of astronomical sources,  combining high sensitivity, low background and medium resolution at FUV wavelengths.  New observational modes have expanded the bandpass beyond the original design limits and reopened wavelength regimes not accessible since the end of the FUSE mission.  COS is expected to continue remain in operation at least through 2016. 

\acknowledgments
The success of the Cosmic Origins Spectrograph has been made possible through the efforts of countless individuals at the University of Colorado, Ball Aerospace, the Goddard Space Flight Center, the Kennedy Space Center and at the Space Telescope Science Institute, and by the crew of STS 125 during Hubble Servicing Mission 4. We would especially like to acknowledge the dedication of Frank Cepollina, Deputy Associate Director, Hubble Space Telescope Development Office.


\bibliographystyle{spr-mp-nameyear-cnd}

\makeatletter
\let\clear@thebibliography@page=\relax
\makeatother

\end{document}